
\input harvmac

\def \U { {\cal U}}

\def \V { {\cal V}}
\def \E { {\cal E}}
\def \P { \Phi }
\def \S { {\cal S}}

 \def \k1 {{1\over
k}}  \def \ov { \over }

\def \O {\Omega }

\def \ra {\rightarrow}

\def \a {\alpha}
\def \b {\beta}

\def \sh {{\rm sinh \  }}

\def \ln {{\rm \ ln \  }}

\def \ch {{\rm cosh \  }}
\def \th {{\rm tanh \  }}

\def \1p {{1\over  \pi }}
\def \2p {{{1\over  2\pi }}}
\def \4p {{ {1\over 4 \pi }}}
\def \8p {{{1\over 8 \pi }}}
\def \p {\phi}

\def \m {\mu }
\def \n {\nu}

\def \k {\kappa }

\def \fourth {{\textstyle{1\over 4}}}

\def \e#1 {{{\rm e}^{#1}}}
\def \const {{\rm const }}

\def \eq#1 {\eqno {(#1)}}

\def \O {{\cal O}}
\def \hr {{\hat r}}

\def \fourth {{1\over 4}}

\def \e#1 {{{\rm e}^{#1}}}
\def \const {{\rm const }}

\def \vp {\varphi}

\def\np {  Nucl. Phys. }
\def \pl { Phys. Lett. }
\def \mpl { Mod. Phys. Lett. }
\def \prl { Phys. Rev. Lett. }
\def \pr  { Phys. Rev. }


\lref \ortin {T. Ortin, \pr {\bf D47} (1993) 3136.}
\lref \kaloper { N. Kaloper and K.A. Olive, University of Minnesota preprint
UMN-TH-1011, 1991. }
\lref \kallosh {R. Kallosh, A. Linde, T. Ortin, A. Peet and A. Van Proeyen, \pr
{\bf D46} (1992) 447.  }
\lref \derendi {J.P. Derendinger, L.E. Ib\~ anez and H.P. Nilles, \pl {\bf
B155}
(1985) 65. }
\lref \quiros {J. Garcia--Bellido  and M. Quir{\'o}s,  \np {\bf B385} (1992)
558. }
\lref \gibb {G.W. Gibbons and K. Maeda, \np {\bf B298} (1988) 741. }
\lref \ghs {D. Garfinkle, G.T. Horowitz and A. Strominger, \pr {\bf D43} (1991)
3140.  }
\lref \cas {  J.A. Casas, Z. Lalak, C. Mu{\~n}oz and G.G. Ross,
\np {\bf B347} (1990) 243;
 B. de Carlos, J.A. Casas and  C. Mu{\~n}oz,  \np {\bf B399} (1993) 623.}

\lref \horow {G.T. Horowitz, ``The dark side of string theory: black holes and
black strings", in
{\it Proceedings  of the 1992 Trieste Spring School on String Theory and
Quantum Gravity"}, preprint
UCSBTH-92-32. }
 \lref \maison {G. Lavrelashvili and D. Maison, \pl {\bf B295} (1992) 67. }

\lref \bizon {P. Bizon, \pr
{\bf D47} (1993) 1656. }

\lref \bhym {G. Lavrelashvili and D. Maison,  Munich preprint MPI-Ph/92-115;
P.
Bizon, University of Vienna  preprint ESI-1993-18 (1993); E.E. Donets and
 D.V. Galtsov, Moscow University preprint
DTP-MSU-93-01 (1993). }

 \lref \zumin {M.K. Gaillard and B. Zumino, \np {\bf B193} (1981) 221. }
\lref \dine {M. Dine, R. Rohm, N. Seiberg and E. Witten, \pl {\bf B156} (1985)
55. }
\lref \bartnic {R. Bartnik and J. McKinnon, \prl {\bf  61} (1988) 141; D.V.
Galtsov and M.S. Volkov, \pl
{\bf B273} (1991) 255. }
\lref \BrSt{ R. Brustein and P. Steinhardt,  \pl {\bf B302} (1993) 196.   }
\lref\derovr{G. Cardoso and   B. Ovrut, Nucl. Phys. {\bf B369} (1992) 351 ;
J.P.
 Derendinger, S. Ferrara, C. Kounnas and F. Zwirner,
Nucl. Phys. {\bf   B372} (1992) 145; J. Louis, in {\it Proceedings of the
International Symposium on
Particles, Strings and Cosmology}, Boston, March 1991,  P.
Nath and S. Reucroft eds. (World Scientific, 1992).}
\lref \cadoni {M. Cadoni and S. Mignemi, Cagliari Univ. preprint INFN-CA-10-93.
 }
\lref \tse { A.A. Tseytlin, ``String cosmology and dilaton",  in {\it
Proceedings of the 1992 Erice
workshop
 ``String Quantum Gravity  and Physics at the Planck scale"}, ed. N. Sanchez
(World
Scientific,1993).}
 \lref \stewart { N.R. Stewart, \mpl {\bf A7} (1992) 983.}
\lref \HH {J.H. Horne  and G.T. Horowitz,
\np {\bf B399} (1993) 169.}
\lref \GH { R. Gregory and J.A. Harvey, \pr {\bf D47} (1993) 2411. }
\lref \witten { E. Witten, \pl {\bf  B155} (1985) 151.
}
\lref\kap{V. Kaplunovsky, \np {\bf B307} (1988) 145.}
\lref \DKL {L. Dixon, V. Kaplunovsky and J. Louis, \np {\bf B355} (1991) 649.
}
\lref \shapere { A. Shapere, S. Trivedi and F. Wilczek, \mpl {\bf A6} (1991)
2677. }
\lref \font { A. Font, L. Ib\~anez, D. L\"ust and F. Quevedo, \pl {\bf B249}
(1990) 35;
 A. Sen,  Tata Institute preprint TIFR-TH-92-41 (1992); J. Schwarz,
 Caltech preprint CALT-68-1815 (1992);
J. Schwarz and A. Sen,  Santa Barbara Institute 
 preprint NSF-ITP-93-46 (1993).}
\lref \ffont { A. Font, L. Ib\~anez, D. L\"ust and F. Quevedo, \pl {\bf B245}
(1990) 401;
\np {\bf B361} (1991) 194.}
\lref \fer {S. Ferrara, D. L\"ust, A. Shapere and S. Theisen, \pl {\bf  B225}
(1989) 363. }
\lref \anton {I. Antoniadis, J. Rizos and K. Tamvakis, Palaiseau
preprint CPTH-A239.0593 (1993). }
\lref \IbLu   { L. Ib\~ anez and D. L\" ust, \np {\bf B382  } (1992) 305.    }
\lref\CFILQ{M. Cveti\v c, A. Font, L. Ib\~anez, D. L\" ust and F. Quevedo
 \np {\bf B361} (1991) 194.}
\lref\shenker{S. Shenker, in {\it Proceedings of the Carg\` ese Workshop on
Random Surfaces, Quantum Gravity and Strings} (1990).}
\lref\brov{ R. Brustein and
B. Ovrut, Univ. of Pennsylvania preprint, UPR-523-T (1992).}
\lref\cqr{ M. Cveti\v c, F. Quevedo
and S.-J. Rey, \prl {\bf 63} (1991) 1836; E. Abraham  and P.
Townsend,  \np {\bf B351} (1991) 313;
M. Cveti\v c, S. Griffies and S.-J. Rey, \np {\bf   B381}  (1992) 301.}
\lref\khuri{  R. Khuri, \pr {\bf D46} (1992) 4526;
 J. Gauntlett, J. Harvey and J.T. Liu, University of Chicago preprint
EFI-92-67 (1992); M. Duff and R. Khuri, Texas A\& M preprint,  CTP-TAMU-17/93
(1993).}

\lref \ferr {S. Ferrara, N. Magnoli, T. Taylor and G. Veneziano, \pl {\bf B}245
(1990) 409.}
\lref \nill {P. Nilles and M. Olechowski, \pl {\bf B}248 (1990) 268. }
\lref\cvts{M. Cveti\v c and A.A. Tseytlin,  preprint CERN-TH.6911/93;
hep-th/9307123.}


\baselineskip14pt

\Title {\vbox {\baselineskip14pt\hbox{Imperial/TP/93-94/15}
 \hbox{hep-th/9402082}}}
{\vbox{\centerline {String    solutions with non-constant scalar fields }
}}
\centerline{ A.A. Tseytlin\footnote{$^{*}$}{
On leave from P.N. Lebedev Physics Institute, Moscow, Russia } }
\vskip2pt
\centerline{\it  Theoretical Physics Group }
\centerline {\it  Blackett Laboratory}
\centerline{\it Imperial College}
\centerline{\it  London SW7 2BZ, United Kingdom }

\bigskip\bigskip
\centerline {\bf Abstract}
\medskip
\baselineskip10pt
We discuss   charged  string
solutions  of the effective equations  of
 the   $D=4$  heterotic string theory
 with non-constant  dilaton  $\p$ {\it and}
 modulus $\vp$ fields.  The effective action contains
the  generic
moduli-dependent  coupling function   in
 the gauge field  kinetic term and non-perturbative scalar
potential.

\vskip100pt
\noindent
 To be published in   $\ $   {\it    Proceedings of 27-th
 International Symposium  on Theory of Elementary Particles,
Wendisch - Rietz, September 1993.}

\Date
 {January  1994}

\vfill\eject

\noblackbox
\baselineskip 16pt plus 2pt minus 2pt
\nopagenumbers

\newsec{Introduction}
Scalar fields play a very important role in string theory.
They are related to the two basic infra-red problems:
why there is a correspondence with the Einstein theory
(why there is no extra scalar mode of gravity
in spite of the existence of the  massless universally coupled dilaton in the
free string spectrum)
and why the cosmological constant should be zero.
Both the dilaton and moduli (string modes   associated with
 compact extra dimensions) are natural partners of the metric
and thus should play an important role in string gravitational physics.

Scalar  fields  are known to couple   to the  kinetic terms of the gauge
fields.  In particular,
in string theory,  the dilaton field determines the strength  of the gauge
couplings at the tree level of the effective action, while  string one-loop
(genus-one) contributions   give  moduli dependent corrections to such
couplings.
Thus, in general,   a scalar function  $f$ that couples to  the gauge field
kinetic energy is  a function of both the dilaton as well as the moduli.

The dilaton and  the moduli have no potential in the effective action  to all
orders in
string loops.  To avoid a  contradiction with observations
they should  acquire masses.  Currently proposed scenarios
rely on  non-perturbatively  induced potential $V$ due to gaugino condensation
in the hidden gauge group sector.
 Such a potential would  generate masses
for the dilaton and the moduli
 and  at the same time provide a  mechanism   of  supersymmetry breaking.
 While these  scalar fields    eventually get masses,
  they may change with distance (or time, in the cosmological context) at small
scales (or times),
i.e.   may participate in the dynamics at
 small scales (or times),
 { or} in the region  where  non-perturbative effects presumably can be
neglected.
String solutions are usually   discussed in
perturbation theory in $\a'$.
For example,  new   charged black hole  string  solutions \gibb\ghs\ have been
recently obtained   by
taking into  account  the  tree level coupling of the dilaton to the
 gauge fields (for a review see \horow).
Below we shall consider (following \cvts) charged string solutions
taking into account
perturbative (genus-one,  moduli dependent threshold corrections to the
gauge couplings),  and  non-perturbative  string effects
 (non-perturbatively induced potential
for the dilaton and moduli).
\foot{Aspects of charged dilatonic black hole solutions with non-perturbatively
induced dilaton
mass included
 were addressed  in \GH\HH.
 However,
 the potentials for the dilaton were  not always taken
 to be `realistic' or well motivated from the point of view of
 non-perturbative dynamics like gaugino condensation in the hidden gauge
 sector of
the gauge group.}
Such corrections may substantially modify the tree level solutions and
it  is thus important to include
them in order to understand  predictions of string theory.
 We shall
consider      abelian electric  and magnetic solutions
 in flat space concentrating on the role of the non-trivial functions
$f$, a  coupling function of scalars  to the gauge field kinetic  term, and
 $V$, a non-perturbative potential for the dilaton and the moduli.

The solutions we shall describe \cvts\    should have generalisations to the
curved space.
We shall assume that they approximate the exact solutions  of the whole set
of equations (including the gravitational one) in the region where the
curvature is small.
As in the case of solitonic solutions in  field theory in flat space  one can
ignore
graviational effects  if the  scale of the solutions is  large  compared to the
 gravitational scale $\sim E/M_{Pl}^2$
(i.e.  if the  energy of the solutions is small enough).
It is true that in the absense of  a non-perturbative potential the dilaton and
the metric
are on an equal footing. Once the potential is generated, it introduces a new
scale
(different from the Planck one). This   makes  it possible  in principle
to  `disentangle' the metric from the dilaton  and to consider  dilatonic
solutions
in flat space  with a characteristic scale  being larger than   the
graviational one.

We shall
include the dilaton $\p$  and only one    modulus field $\vp$   (which is
associated with  an overall compactification scale)
and look  for stable spherically symmetric
finite-energy  solutions  with  a  regular  gauge  field
strength in flat $D=4$ space-time.
We shall  consider a
general class of functions $f$ and $V$.  We shall treat examples
with $f$ modified by the string loop corrections, as they appear \DKL\ in a
class of orbifold-type  compactifications, and with the
non-perturbative  potential
$V$  due to the gaugino condensation in the hidden sector of the theory,
as special cases.
 We shall ignore  higher derivative terms, assuming that  the
fields change slowly in space.

It turns out that the abelian electric solutions are regular, have finite
energy,
and  are stable when the abelian subgroup is
 embedded in a non-abelian gauge group.
They  have the  effective string coupling $\e{\p} $  increasing from  zero
at the origin ($r=0$) to a finite value $\e{\p_0} $ at $r=\infty$. The
asymptotic value  $\p_0$ of the dilaton
corresponds to the minimum of the potential $V$.
Thus the small  distance region is a weak coupling region  and
can be studied ignoring non-perturbative corrections. The   large distance
region
corresponds to  the `observed' world where the dilaton is trapped in the
minimum
of $V$.
 The modulus field $\vp$ is slowly  varying with $r$; at large scales it is
fixed
at  the minimum of the potential $V$, while at small scales its value
decreases slightly. Generic  existence of such  `non-topological' solitonic
  charged    configurations   may have
potentially interesting applications.
As for  abelian magnetic solutions,
 here at small distances   ($r\ra 0$) the  dilaton  approaches
the  strong coupling region, $\e{\p} \ra  \infty$, while
 the modulus  goes to zero, $\vp\ra 0$, {i.e.},
  the small scale region is the compactification region.
The role of the non-perturbative potential is again to fix the asymptotic
values of $\p_0$ and $\vp_0$ to be at its  minimum.

\newsec {Low-energy string effective equations}

The low-energy $D=4$ effective action of the heterotic string theory
has the following structure
$$S=
{1 \ov 16 \pi }
\int d^4 x {\sqrt {-g }}\ \left[ R - 2 \del_\mu \p\del^\mu \p - 2 \del_\mu
\vp\del^\mu\vp - f(\p, \vp) F_{\m \n} F^{\m\n}
 -  4 V (\p , \vp) + ...  \right] \ \ . \eq{2.1} $$
 For simplicity we consider,   along with
the dilaton field $\p$,  only
 one modulus field $\vp$, associated
 with an overall scale of compactification and  ignore he axionic
partners  of $\p$ and $\vp$ as well as other matter fields.
At the tree level
$$ f_{tree}= \e{-2\p} \ \ , \ \ \  \ \  V_{tree}=\  c \ \e{2\p} = 0 \ .
\eq{2.2} $$
In the case  of the supersymmetric $D=4$ heterotic string
$V$ remains zero to all orders in the string
perturbation theory. On the other hand, the gauge coupling function $f$
receives a non-trivial,
$\vp$-dependent, string one-loop  (genus-one) correction \DKL
$$ f_{perturb} = \e{-2\p} + f_2(\vp) \  , \ \ \ \  V_{perturb} =0 \ . \eq{2.3}
$$
The modulus dependent function
$f_2$  depends on a type of superstring vacuum one is considering. In
particular, for toroidal compactifications and  a class of orbifolds, it
is invariant under the duality  symmetry ($\vp \to -\vp$) and
 can be schematically written in the form:
 $$ f_2(\vp)  = b_0 \ln\left[(T + T^*) |\eta (T)|^4 \right] +b_1,\ \ \
T=\e{2\vp/\sqrt 3} \ \  .
\eq{2.4} $$
It turns out that  $\ln  \left[(T + T^*) |\eta (T)|^4\right]$
is always negative,  has a maximum at $\vp=0\,(T=1)$ and approaches
$ - {\pi\ov 3}T =-{\pi\ov 3}\e{2\vp/\sqrt 3}
$
 as  $\vp \ra \infty$. The constant $b_0$
is related to the one-loop $\b$-function coefficients associated with the $N=2$
subsector of the massless
spectrum in a symmetric orbifold compactification.
Generically
$b_0=\O (1/100)$  and  is negative (positive) in the case of  the abelian
(non-abelian)
 gauge  group factors.

Since the string  coupling is related to the dilaton,
both $f$ and $V$  may contain non-perturbative contributions
which are  non-trivial functions of $\p$.
Such terms in $f$ do not actually appear in
the proposed gaugino condensation scenario for
 supersymmetry breaking. In general, however, one should allow for a
possibility that
 $f$ may contain additional, non-perturbative
terms which depend on $\p$, {\it e.g.},   $\exp{[- k
\exp (-2\p)]} $ (implying   $ 1/g^2 \ra 1/g^2 \  + \ a \  \e{ - k/g^2}  $).
A detailed  structure of a non-perturbative potential $V(\p, \vp)$
depends  on a  particular mechanism of
supersymmetry breaking.  We shall  describe the form of
$V$ due to the gaugino condensation in the hidden  sector
of the  gauge group \derendi\dine\ffont\ferr\nill.

In the case of  symmetric  orbifolds  with  the
compactification moduli
of all three two-tori equal to $T$ the  one finds \ffont\ferr\nill
$$   V_{non-perturb.}\equiv V(S,T)
 ={|H|^2 \ov |\eta (T)|^{12} S_RT^3_R } \left[ |S_R {{\del \ln H} \ov \del S }-
1  |^2
 + {3\ov 4\pi^2}     T^2_R    |{\hat G}_2(T)|^2 -3   \right] \ , \eq{2.5} $$
where
$$ H(S) =
\sum^J_{i=1} d_i \e{-a_i S}  \ , \  \ \ a_i= {3\ov 2b_{0i}}\ ,
\eq{2.6} $$
 $J$ is a number of gaugino condensates and $b_{0i}$ are
the (one-loop $N=1$)
$\b$-functions
of the gauge group factors  of the hidden gauge group sector.  Also,
$S_R = 2 {\rm Re} S ,  \ T_R= 2 {\rm Re} T $ ,  and
 $$\ {\hat G}_2(T) =  G_2 (T) -2\pi T^{-1}_R=  - 4 \pi {1\ov \eta (T)} {\del
\eta (T)  \ov \del
T } - 2\pi T^{-1}_R \ .   $$
This potential  vanishes in the weak
coupling limit $S = \e{-2\p} \ra \infty $ and has an extremum in
$S$ if  $ S_R {\del W\ov \del S }- W
=0$.  A minimum exists if $J >1$, {i.e.},
in cases with more than one gaugino
condensate.
 As for the extrema in $T$,   $\del V/ \del T = 0$, they are achieved
  at the self-dual points
  $T=1$ and
$T={\rm e}^{i\pi /6}$,
which are saddle points of $V$, and at $T\sim 1.2$, which is
the  minimum of $V$.
At  a fixed,  extremal value of $T$  and  a fixed ${\rm Im} S$
the potential $V$ can be represented in the following form
$$ V= S_R^{-1} \sum^J_{i=1} {\rm e}^{-a_i S_R} (c_i  + d_i S_R + e_i S_R^2) \ ,
 \eq{2.7} $$ where $a_i$, $c_i$, $d_i$, and $e_i$ are constants
and $S_R = 2 {\rm e}^{-2\p}$.
  For example, in  the case of two gaugino condensates,
     $J=2$,  this
potential starts from zero in the  weak coupling region $\p \ra -\infty$,
grows  and reaches a local
maximum, then  decreases to a local minimum (with negative value of $V$),
then  has the second  local
maximum and finally  goes to $-\infty$ at   $\p \ra +\infty$. Since the
potential  has a local minimum in $\phi$, it may
fix the value of the dilaton.\foot{This would
 give  a mass  to the fluctuating part
of the dilaton.
A generic  property of this potential is that starting from a weak coupling
region it first
increases and has a local maximum and only then decreases to a minimum, namely,
the potential is {not} convex everywhere.
 Another problem is that the value of the potential at the
minimum, {i.e.} the effective cosmological constant, is negative in general.
A local minimum with a zero cosmological constant can be achieved in the case
 with
  more than two gaugino condensates.
Note, however, that in this case there
are usually also other minima with negative cosmological constants (see {\it
e.g.}, \BrSt).}

 We shall discuss
 charged solutions with  non-trivial
dilaton $\p$  and  the modulus field $\vp$
in flat four dimensional  ($D=4$)  space-time, {i.e.},
$ds^2=-dt^2 + dr^2 + r^2
d\Omega^2$. The  relevant part of the action (2.1)  can be put in the form
  $$S= {1\ov  4   \pi } \int d^4 x \
\{  - \ha   (\del \P_i)^2 -  \fourth f(\P_i) F_{\m \n} F^{\m\n}
 -  V (\P_i) \} \ \ ,  \eq{2.8} $$
so that the corresponding  field equations  are
 $$ D_\m (f F^{\m\n}) =0 \ ,\ \ \  D_\m  F^{*\m\n}
=0 \ , \eq{2.9} $$ $$ D^2 \P_i -  \fourth  \del_i f  F_{\m\n} F^{\m\n} - \del_i
V = 0 \ , \ \ \ \
i=1,2  \ , \eq{2.10} $$
 where  $ \del_i= {\del/\del \P_i} $ and $ \P_i = (\p , \vp )
$. It is easy to see that this system  transforms into the same  one  under
 $ f\ra  f^{-1} \  , \ \ \
  F_{\m\n} \ra f  F^*_{\m\n} \ , \ \  \P_i\ra\P_i \ , $  so
 that  electric solutions  for the action (2.8)  with
the  gauge coupling function $f(\P_i)  $ are   related to the magnetic
solutions of the  theory with
the coupling  $f^{-1}(\P_i) $.

 Let us consider  first the electric solution:
$$  F_{01}= E(r) \ ,  \ \  \ F_{0k}=F_{1k}=F_{kl}=0\ (k,l=2,3) \ , \ \ \ \
 \P_i = \P_i (r) \ . \eq{2.11}
$$
Then
$$  {d\ov dr} ( r^2 f E) = 0 \  ,
\ \ \  E=  {q\over { r^2 f(\P_i (r))}} \ ,  \eq{2.12} $$
so that   (2.10) becomes ($i=1,2$)
$$ \P_i'' +  \del_i U -  {1\ov x^4 } \del_i V = 0   \ ,  \ \ \ \ \ U\equiv
-{q^2\ov {2f}} \ .
 \eq{2.13} $$
We have introduced the  new coordinate
$ x = {1/ r} \ $ with the range  $  0\le x \le \infty $,  and
$\P_i'\equiv d \P_i / dx  $.
Equation  (2.13) can be interpreted as corresponding to a  mechanical system
with the action
which,  at  the same time, gives the energy  of
 the  field configuration  as  derived from  (2.8):
$$ \E = \S =
 \int^\infty_0  dr \ r^2  \ [ \
 \ha  ({d\ov dr} \P_i)^2
 +   {q^2\ov  2r^4  f }
  +   V \ ] =
\int^\infty_0  dx  ( \ha  \P_i'^2
 - U
 + {1\ov x^4 }  V ) \ .  \eq{2.14} $$
The case with $V=0$ corresponds to the mechanical system with the
conservative potential $U$, while the case with $V\neq 0$
corresponds to the mechanical system with   non-conservative
(time-dependent) potential
$$\U (\P_i, x)\equiv  U -  {1\ov x^4 } V \ . \eq{2.15} $$

One can show that a sufficient  condition of
 `linearized' stability  of  the abelian  electric solutions
is
$$\V_{ij}\ge 0 \ , \ \ \   \V_{ij}  \equiv  - \del_i \del_j U   +
r^4 \del_i \del_j  V  = -  \del_i \del_j \U  \ . \eq{2.16} $$
 This condition has  an obvious interpretation that
perturbations should not decrease the energy (2.14) of the system.

The set of equations, the energy  and the condition of stability for the
magnetic  solution
 $$F_{23}= h  \sin \theta   \ ,  \ \  \  \  h= {\rm const.} \  , \ \ \
 F_{01}=F_{0k}=F_{1k}=0\  , \ \
 k=2,3 \ ,  \ \  \ \ \P_i = \P_i (r) \ ,  \eq{2.17}  $$
are found  by replacing $f$ by $f^{-1}$  and $q$ by $h$ in the
above equations or  by using
 $$ U\equiv -{1\ov 2} {h^2f}\ \ . \eq{2.18} $$

\newsec {Solutions in the case of zero scalar potential }
The
properties of the solutions  are different depending     on whether $V$ is
zero or not: in  the former case the system is conservative, in the
latter it is not.    Here we  shall consider
  the case  when the effect
of the potential can be ignored, {i.e.},   when non-perturbative
corrections are small.
 In addition,  we shall start with the casewhen only one
scalar  field  $\P$   is
changing  with $r$.
Such a reduction to only one  field
 could be  possible if  $f$  had extrema with respect to
the other field and thus the other field
 could  be `frozen', {i.e.}, put to a fixed $r$-independent value.
\foot{ In principle, $\P$  may be either the dilaton or  the  modulus,
but  it does not seem to be possible  to
`freeze out' the $r$ dependence of the dilaton in general.}

 If $V=0$   eq. (3.6)   has one integral of motion , {i.e.}, the
`energy' of the corresponding mechanical system:
$$    \ha  \P'^2  +  U(\P) =
 C=\const  , \eq{3.1} $$
where $U=-{1\ov 2}q^2f^{-1}$ and $U=-{1\ov 2} h^2f$ for the electric
 and the magnetic solutions, respectively. We shall assume  that  $f$ is always
positive (since it is
the coupling function in the gauge field kinetic term) and thus,
$U$ is  always negative.
  In looking for  minima
of $\E$ it is useful to represent
 (2.14)  in the form
$$ \E =
\int^\infty_0  dx  \  \ha  \left( \P' \pm
 \sqrt {2C-2U(\P)}\right) ^2 \mp
\int^{\P_\infty}_{\P_0} d\P\sqrt{2C-2U(\P)}-
\int_0^\infty dx C  \ ,   \eq{3.2}
$$
where ${\P_\infty}\equiv\P(x=\infty)$ and $
 {\P_0}\equiv\P (x=0),  \  $ $x\equiv {1/ r}$.
 We shall consider solutions  with a regular field value,
{i.e.}, $\P_0\ne\infty$
 at $x=0$. However, in the core of the solution, {i.e.}, at $x\ra\infty$
the field may or may not blow up.

 The energy is
minimized  if (cf. Bogomol'nyi conditions) $$  \P' =\mp\sqrt{2C-2U(\P)}\  .
\eq{3.3} $$
  The constant $C$ is chosen such
that the  second and the third term in (3.2) give a finite value of the
energy.
Eq. (3.3) implies  the explicit form of solution
$$\int^{\P (x) }_{\P_0}
d\P {1\ov {\sqrt{2C-2U(\P)}}}   = \mp  x \ . \eq{3.4}   $$
 The upper (lower) sign solutions  correspond to $\P$
decreasing  (increasing) with increasing
$x={1/ r}$.
Then  the  energy $\E$, the charge  $Q$
and  the  scalar charge $D$  of the  electric solution
are   $$  \E =
\mp  \int^{\P_\infty}_{\P_0}d\P \sqrt{2C-2U(\P)}
-\int_0^\infty dx C\ ,  $$
  $$Q= [r^2 E(r)]_{r\ra \infty}  = {q\ov f(\P_0)} ,
\ \ \
  D = -[r^2 {d \P \ov dr }]_{r\ra \infty}=
\mp
\sqrt{2C+{q^2\ov f(\P_0)}} \ . \eq{3.5} $$
The magnetic solution  is found from the electric solution
by the replacements $f\ra f^{-1}$ and $q\ra h$, or in other words, by taking
$U=-{1\ov 2} h^2f$.

In order to  have  a finite energy,
 regular solution the potential
 $U(\P)=-{1\ov2}{q^2 f^{-1}(\P)}$ (for the electric solution) and
 $U(\P)=-{1\ov2}{h^2f(\P)}$
 (for the magnetic solution) should satisfy
certain conditions.
The nature of the solution is different in the cases with $C=0$ or $C\ne 0$.
 One can show that the regular, finite energy  solutions
 with the
upper sign, exist only for  the choice of
$C=0$. Such solutions have the  property
$\P_\infty\ne\infty$ and  as $\P\rightarrow \P_\infty$,
  $U(\P)$ approaches zero
 faster than $   (\P-\P_\infty)^2$.  Solutions of this type
correspond to    the case   of  `dilatonic'-type electric solutions
and a
class of `moduli'-type  magnetic solutions.
  Solutions with the lower sign exist for
$C=0$ or $C\ne 0$, depending  on the
nature of  $U$. If $C=0$ one obtains
$\P_\infty= \infty$ and as  $\P\rightarrow \infty$,
 $U\rightarrow 0 $ faster than  $\P^{-2}$ .
 We shall see that such properties are found for a class of
`moduli'-type electric solutions and  a special case of the `dilatonic'-type
 magnetic solution.
On the other hand, there also exist  regular, finite energy solutions
for  a specific value  of $C=C_0> 0$; this is the case  if   $\P_\infty=\infty$
and  as $\P\rightarrow\infty\ $,
  $U(\P)-{1\ov2}C_0$  approaches  zero faster than $\P^{-1}$.

The simplest example  of the dilatonic solution  corresponds to the case of the
tree level dilaton
coupling  $$  f = \e{-2\P} \ , \ \ \ \ \ \P=\p   \ . \
\eq{3.6}  $$
The solution is given by
 eq. (3.5) with  the upper sign  and   $C=0$ (in order to avoid a singularity
at finite $x$). The explicit form of the solution, its  mass $M$,   charge $Q$
 and the
scalar charge $D$  are
$$\p= \p_0 - \ln ( 1 +  {M\ov r})  \  , \  \ \ \
M\equiv \E=  |q| {\rm e}^{\p_0},\eq{3.7} $$
$$ \  E(r) = { Q \ov (r + M )^2 } \ \ , \  Q= M{\rm e}^{\p_0} \  , \ \
D = - M \ .  \eq{3.8} $$   The string coupling
grows from zero at $r=0$ to a finite value ${\rm e}^{\p_0}$ at large distances.
The small distance region is thus a
 { weak coupling} region. Therefore it is consistent to ignore
 the non-perturbative potential $V$ in this region.
We shall see that once the potential $V$
is included, $\p$ will be evolving to its minimum at large $r$.
The   solution   is  stable  since the condition  (2.16)
 is satisfied for (3.6).

The  regular  magnetic solution with the finite energy, a counterpart
of the electric solution
(3.7), is obtained  from the lower sign solution of eq. (3.4)  where
 $f$ in (3.6) is replaced  with   $f^{-1} $, and $q$ with $h$, and
  $C=0$. Then
$$\p=\p_0+\ln (1+{M\over r})\ ,\ \  \E\equiv M=|h|{\rm e}^{-\p_0}\ ,\ \  D=M \
. \ \
\eq{3.9} $$
  The small distance  region  of the magnetic solution  is a
strong coupling region
and  hence  there
non-perturbative corrections (inducing  $V\neq 0$) can be significant.
 Like the corresponding regular electric solution, this
magnetic  solution is also stable.

The above expressions for the dilaton in the electric
  and  the magnetic  solutions
coincide  with the expressions for the dilaton  in the electric and
magnetic black hole solutions of \gibb\ghs\  if  $r$ is identified with the
coordinate $\hr
 =r -  M$, where $r$ is defined outside the horizon.  In terms of $r$
we get
asymptotic    large $r$ expressions for the dilaton  of \gibb\ghs.
For example,  in the case of the
electric  black hole  solution  the  metric (in
the  Einstein frame) takes the form (see, {e.g.}, \horow)
 $$
 ds^2 = -(1- {m\ov \hr})(1 + {M \ov \hr})^{-2} dt^2 +
(1- {m\ov \hr})^{-1} d\hr^2 + \hr^2 d\Omega^2
\ , $$
while the expressions for the dilaton and the electric field  coincide with
 (3.7) and (3.8)
where $r$ is
replaced by  $\hat r$.
The physical mass is $\m = M + m$ and $Q=(M\m)^{1/2}$.
As was discussed  in \horow\ the small $\hr$ region is a weak coupling
 region for the electric
solution but a strong coupling region for the magnetic one (which is obtained
from the electric
 solution by the duality transformation $\p \ra -\p, \ F_{\m\n} \ra {\rm
e}^{-2\p}  F^*_{\m\n}
$).

An  important  generalization of  (3.6)
corresponds to
$$  f = \e{-2\P} +   b \   \ , \  \ \ \ \P=\p \ ,
\eq{3.10}  $$
where the constant $b$  can be interpreted, { e.g.},  as
 a contribution of  threshold corrections to $f$
 in the case when the  space dependence
of the modulus   field can be ignored.    Assuming  that $b >0$, we get the
electric solution
from  eqs.(3.4),(3.5)   with the choice of the upper sign and $C=0$:
$$  [-\sqrt { \e{-2\p}  +   b}
+ {\sqrt b} {\rm Arcsinh} ( {\sqrt b}\e{\p} )]^{\p(r)}_{\p_0}  = -{|q|\ov r }
\ ,   \ \ \ \E  =
{|q|\ov \sqrt b} {\rm Arcsinh} ( {\sqrt b}\e{\p_0} )
 \ .   \eq{3.11} $$
 Here $\p_\infty = -\infty$,
{i.e.},  the string coupling
  is increasing with $r$  from zero to a finite value $\e{ \p_0} $
and the energy is finite.
The stability condition (2.16) is satisfied  if $\p_0$ is such that
$\e{-2\p_0} -   b \geq 0$.

 It turns
out  that the only
finite energy, regular magnetic  solution exists for  $b<0$ and the choice of
$C=-h^2 b>0$. In
this case:
$${1\ov \sqrt{-b}} [{\rm Arcsinh} ( \sqrt{-b}\e{\p} )]^{\p(r)}_{\p_0}
=  {|h|\ov r } \ ,  \ \ \ \E= |h| (\sqrt { \e{-2\p_0}    -b}-\sqrt{-b})
\ .  \eq{3.12} $$
For  $b>0$  one  gets  a  regular solution  with
$\p_\infty = \infty $
 but infinite energy.

Next, let us consider `modulus-type'  solutions   corresponding to
$$ f(\P) = p^2( \ch  a\P +  s )^2 \ . \eq{3.13} $$
 For large negative $\P$ and $a=1$  this function  is the same as in (3.6).
 Being  symmetric under the `duality' symmetry   $\P \ra - \P$
this $f$ (with  $\P = \vp$, $a= 1/\sqrt 3$)  is a good approximation for
  the
modular invariant  coupling function $f_2(\vp)$
in (2.4).
 A non-zero  constant $s$ may  be
considered as accounting for  a modulus independent contribution to  $f$.
Without loss of generality one can  fix the boundary condition
$\vp_0\equiv \vp(1/r=0)>0$
 (due to the duality symmetry $\vp\to-\vp$  solutions
with  the boundary condition
 $\vp_0<0$ are related to the ones with $\vp_0>0$).
 Then  for the regular, positive energy
 electric solutions   one finds from eqs.(3.4),(3.5)  (with the
 {lower} sign and
$C=0$)
$$[ \sh a\vp  +  sa\vp]^{\vp(r)}_{\vp_0} =   {|q|a\ov |p|r}    \
 \ ,   \eq{3.14} $$
$$\E
={ 2|q|\ov a|p|\sqrt{1-s^2}} [{\rm Arctan }({\sqrt{1-s^2}\ov 1+s }
\th {a\vp\ov 2}) ]^{\vp_\infty}_{\vp_0} \  , \ \ \ s^2<1   \ ,
\eq{3.15}
$$
$$ E=  { q\ov  r^2 p^2[ \ch  a\vp (r)  +  s ]^2  } \ ,\ \
 Q=  {q\ov p^2( \ch  a\vp_0 +  s )^2}\ ,  \ \
  D =
{q\ov  |p|( \ch  a\vp_0 +  s )} \ . \eq{3.16}  $$
 For  $s>-1$ the solution is regular for {any} $\vp_0>0$.
On the other hand, for $s<-1$,  the  regular solution  exists when  $\vp_0>0$
satisfies $ \ch  a\vp_0 +  s  > 0 $.
 The condition of stability (3.12), {i.e.},
$ {d^2 \ov d \vp^2} \ {f }^{-1}   \geq 0  $,
is satisfied if  $\vp_0$ is such that
$ 2 {\rm cosh}^2 a\vp\  - \ s\ \ch a\vp \ - 3\  \geq 0  $
for all values of $ \vp (r)  $.
Since  there always exists a choice of $\vp_0$ for which both
of the above constraints  are  satisfied  we get a class of stable, regular
finite energy electric  solutions.

In   general, the electric solution  (3.14)--(3.16) has
 the property that it
increases from a positive finite  value $\vp_0\equiv \vp (1/r=0)>0$ to
$\vp_\infty\equiv\vp(1/r=\infty)=+\infty$. Namely, in the core of the
solution, {i.e.}, as $r\rightarrow 0$, a
 decompactification ($\vp\rightarrow
\infty$)
takes place.  There is an obvious  analogy with the corresponding  `dilatonic'
electric
solutions with $f$ in (3.10):  the role of
the weak coupling  at the core of  the `dilatonic' electric solution  is
now   played  by the
decompactification  at the core of the `modulus' electric solution.

We thus see  that  in the case of the    duality invariant
function $f(\vp)$  (3.13) there are  stable, regular,  finite energy
electric solutions.
It turns out  that there are  no regular, finite
 energy magnetic solutions
corresponding to  such $f(\vp)$, unless  $s =-1$ \cvts.  In the latter case
$$  \coth {a\vp(r) \ov 2} -\coth {a\vp_0 \ov2} =  {|p||h|a\ov r} \ ,
 \ \ \ \E= {|h||p|\ov a} (\sh a\vp_0  - a \vp_0 ) \ .
  \eq{3.17} $$
Here  $\vp_\infty
=0$, {i.e.} the magnetic solution corresponds
 to the compactification at the self-dual point $\vp=0$.
 For the duality invariant $f$  with $f(0)=0$  the electric  and the
magnetic  solutions have  complementary features, similar to the ones of the
dilatonic solution with $f$ in (3.6). Now,  however, the role   of the
strong-weak coupling  regions is
played by the compactification - decompactification regions.
One can show \cvts\  that the finite energy electric { and}
 magnetic  solutions with
 qualitatively  the same behaviour  exist
 for a general positive definite, duality invariant function
$f$ with the following properties: $f(\vp)$ has the minimum at $\vp=0$,
$f(0)=0$, and as $\vp\rightarrow\infty$, $f$ grows faster than $\vp^2$.

Let us now  turn to    the solutions in  a more `realistic'  case
when both  the dilaton and the modulus field can change in space.
We shall choose  the  coupling function in the following  form
$$  f(\p,\vp) =  f_1(\p) + f_2 (\vp)  \ , \eq{3.18} $$
which is  a generalization of  the  perturbative expressions (2.3),(2.4):
$f_1= \e{-2\p} $  and  $f_2=
b_0 \ln [(T+T^*) |\eta(T)|^4]+b_1$,  $\ T= {\rm e}^{2\vp/\sqrt 3}$.
The system of equations  for the two scalars $\P_i=(\p,\vp) $  in the
case of the electric solution becomes
$$  \P_i''   + {q^2 \ov 2 (f_1+f_2)^2} {d f_i\ov d\P_i} = 0   \ ,  \ \ \ \
i=1,2\ .  \eq{3.19} $$
It
reduces to the one-scalar case  considered in the
previous section  if   one of the scalars  $\P_i$ is
fixed  to be at the extremum of the corresponding  function $f_i$.
 While  the  tree-level dilatonic
coupling $f_1$  does not have a local extremum
(and thus the dilaton cannot be `frozen'
at a constant value) the  modulus coupling  $f_2$
does have  an extremum  at $\vp=0$.

It is possible to  find   electric solutions  of eq.
 (3.19)  using a  perturbative approach, {i.e.}
assuming  that $f_1 \gg f_2$ \cvts. This assumption is satisfied, in fact,
 in the case
of the threshold correction in  (2.4).
In  such a case one can  reduce the
system  of the two  second-order coupled differential equations (3.19) to a
set of { two first-order} coupled differential equations.
For example,  one can approximate  the function $f_2$ in (2.4) by
 a
simple   duality invariant function  $f_2=p^2\sinh^2a\vp$.   Then
$$ \ln [{\th a\vp (r)  \ov \th a\vp_0} ] = {1\ov 6} a^2  p^2 \e{2\p_0}  \left[
 1- { (1+ {|q|\e{\p_0}
\ov r} )^{-2}} \right] \
.   \eq{3.20} $$
 As $r\rightarrow 0$, $\vp$  increases (decreases) for  $p^2>0$ (
$p^2<0$).
One can show that
the   solutions  with $f_2 >0$  (abelian case) are  unstable
 (since the tree-level dilaton
solution is stable, the instability
 is due to   the modulus sector),
 while those with  $f_2<0$ (embedding in a non-abelian gauge group) are stable.

\newsec{ Case of non-vanishing  scalar potential}
When $V\not=0$ the system  of equations (2.13)
$$ \p'' + {\del U \ov \del
\p} - {1\ov x^4 } {\del V \ov \del \p} = 0   \ ,  \ \ \ \vp''
 +{\del U \ov \del \vp} - {1\ov x^4 } {\del V \ov \del \vp} = 0   \ ,
 \eq{4.1} $$
(where  $U=-{1\ov 2} q^2 f^{-1}$ and  $U=-{1\ov 2} h^2 f$ for the electric and
magnetic   cases)
does
not  have
integrals of motion.
The  non-perturbative potential   due to
 gaugino condensation  in the hidden sector of the gauge group (2.7)   depends
on
both the dilaton,  $S= \e{ -2\p } $  and the modulus, $T=\e{2\vp/\sqrt 3 } $
and  vanishes
 in the limit of
small string coupling $\e{ \p}  \ra 0$.
The potential is not convex
everywhere;  in addition to  the   minimum,  it  also has saddle points and
local
maxima. We  shall consider a class of  functions $f=f_1(\p)+
f_2(\vp)$  (eqs.(2.3),(2.4)).

 Let us consider first the electric solutions. At small
 radius $r\ra 0$ the electric solutions correspond to the weak coupling
region, $\p\ra -\infty$, and thus, the potential $V$ term
can be  neglected in both equations in (4.1) and we are back to the case
discussed in
the previous section.
To   determine  the  large $r$ asymptotic behaviour of
  $\p$ and $\vp$ let us again assume that
 in the large distance  region $\p$ and $\vp$ approach
  constant values $\p_0$ and $\vp_0$.  Using the expansion
$$ \P_i=\P_{0i}  + k_{i}x +l_{i}x^2 + m_{i}x^3 + n_{i}x^4  + ... \  , \ \ \
\P_i= (\p ,\vp)
\  , \eq{4.2}  $$
  one finds that (4.1) is  satisfied  to the leading order  in $x$  if
$$ k_i=l_i=m_i=0, \ \  n_i\ne 0  \ , \ \  \  \  ({\del V \ov \del\P_i})_{\P_0}
=0 \ ,
\ \   \eq{4.3} $$
$$ {q^2\ov 2 }({1\ov f^2 } {\del f \ov \del \P_i})_{\P_0}  -\sum_{j=1}^2\ n_j
({\del^2 V \ov \del \P_i\del \P_j})_{\P_0} =0   \ . \eq{4.4} $$
As a result,
$$ \P_i(r \ra \infty)   = \P_{0 i} + {n_i\ov r^4} +  \O({1\ov r^5}) \ ,
\eq{4.5} $$
 where the  asymptotic value  $\P_{0i}=(\p_0, \vp_0)$   corresponds to an
extremum of $V(\Phi)$. As follows from (2.4),    $f$
 is positive,   ${\del f \ov \del \p} $
is negative and ${\del f \ov \del \vp} $ is positive (negative) for the
abelian  (non-abelian  embedding) case.  For consistency with the behaviour of
the solutions  at
small $r$ (Section 3) we are to assume that  as $r$ increases  $\p$ should be
growing and
$\vp$ should be falling (growing) in  the abelian (non-abelian embedding) case,
 namely,
$n_1<0$, and $n_2>0$ ($n_2<0$). These constraints in turn imply that
in the case  when the  mixed second derivative  term $({\del^2 V\ov \del
\p\del\vp})_{\Phi_0}$
 is small,  the matrix $({\del^2 V \ov \del \P_i\del \P_j})_{\P_0}$ is positive
definite, {\it
i.e.},  both  $\p_0$ {and} $\vp_0$   correspond to the  {minimum } of
$V$.
One can  show  \cvts\ that the potential $V$ in (2.5) has actually
 the  mixed  second derivative equal to zero,
$({\del^2 V\ov \del
\p\del\vp})_{\Phi_0}=0$.

The magnetic  solutions  have  the strong coupling
region $\p\ra +\infty$ at small  radius, $r\ra 0$.   The potential $V$ term,
however,
 {does not}  qualitatively modify  the particular dilaton solution in this
region  \cvts.
 In addition,  the  potential term can be
neglected  in the second equation in (4.1)  since it
is  proportional to $\e{2\p} x^{-4}\propto r^2$.
 Therefore, as $r\ra 0$,  solutions
for the modulus  can be discussed as in Section 3.
Thus, in both  the  electric and the magnetic,
   cases the presence of the  potential $V$  term
in the equations of motion fixes
the asymptotic values of $\p_0$ and $\vp_0$  to be at the    minimum   of $V$.

In the above discussion  we  have  assumed the metric to be flat.
This is consistent  provided  the energy  of the solutions is small enough.
It is important of course to  extend   the analysis  to the gravitational case,
{i.e.} to find the
 the corresponding  black hole - type   solutions  (generalizing those of
\gibb\ghs\GH\HH) of the
string  effective action  with  the
 threshold corrected   coupling $f$ as well as  the  non-perturbative
   potential $V$ included.   Choosing the metric \gibb
$$ ds^2 = - {\rm e}^{2 \nu} dt^2 +
{\rm e}^{4\zeta - 2 \nu} d y^2 +  {\rm e}^{2\zeta - 2 \nu} d\Omega^2 \ ,
\eq{4.6} $$
where  $\nu$ and $ \zeta$ are functions of $y$  one can repsent the system of
equations that follow from
 (2.1) and generalise (2.10)  as corresponding to the following mechanical
system
(together with the zero energy constraint $T+U=0$)
$$ L=T-U\ ,\ \ \  T= \ha \nu'^2 + \ha \p'^2 + \ha \vp'^2 - \ha \zeta'^2\ ,
\eq{4.7}  $$
$$ U= \ha {\rm e}^{2\zeta} - \ha q^2 {\rm e}^{2\nu}  [f_1(\p) + f_2 (\vp)]^{-1}
-  {\rm e}^{4\zeta - 2 \nu} V(\p, \vp)
\ . \eq{4.8} $$
One may find the large distance asymptotics of the fields  which will
generalise  (4.2)--(4.5) and
the results of \HH.

In conclusion,    let us     emphasize   again  the  generic features of the
 stable, finite energy electric
solution.
It is a  particle-like configuration with the weak coupling region  inside the
core.  It  seems  that
 the proper  dilaton  boundary condition  for  analogous  solutions in string
theory
  should be  $\p\ra -\infty$  at  $r\ra 0$, {i.e.},  small string
 coupling at small scales.
This  corresponds to  an appealing  `asymptotic freedom' scenario  in which
 both perturbative and non-perturbative  string corrections  are negligible
 in the small distance region. Thus,
in this region  the tree-level string theory applies, supersymmetry  and other
symmetries are unbroken (in
the cosmological context the small distance region  corresponds to an  early
time era).
On the other hand, the growth of the dilaton with $r$ implies that  at   large
distances, or in `our world',  the string coupling  becomes
  relatively strong,  so that  non-perturbative corrections can no longer
be ignored.  In this region
supersymmetry is spontaneously broken and   both   the dilaton and the modulus
fields are  stabilized
at the minimum of the non-perturbative potential.

\bigskip
 The author  is grateful to M. Cveti\v c
 for the collaboration on the work described in this paper
and would like to  acknowledge also   the support of SERC.
\medskip

\listrefs
\end